# Supersolid behavior in confined geometry

J. T. West, X. Lin, Z. G. Cheng, and M. H. W. Chan

*Dept. of Physics, The Pennsylvania State University, University Park, Pennsylvania 16802*

We have carried out torsional oscillator (TO) and heat capacity (HC) measurements on solid $^4$He samples grown within a geometry which restricts the helium to thin (150 micron) cylindrical discs. In contrast to previously reported values from Rittner and Reppy of 20% non-classical rotational inertia (NCRI) for similar confining dimensions, 0.9% NCRI (consistent with that found in bulk samples and samples imbedded in porous media) was observed in our TO cell. In this confined geometry the heat capacity peak is consistent with that found in bulk solid samples of high crystalline quality.

The observation of mass decoupling below ~200 mK in solid $^4$He under AC oscillation [1,2] has been interpreted as the signal of non-classical rotational inertia (NCRI), putative evidence of the supersolid state [3]. In the same temperature range as the NCRI signals, a heat capacity peak [4] and an increase in the shear modulus of solid $^4$He [5] have also been recently discovered. At slightly higher temperatures and close to the melting curve, evidence of DC flow has been reported [6].

A tremendous amount of work has been done in an attempt to understand the microscopic origin of the NCRI signals [7-9]. These signals have been reproduced in many laboratories [10-16]. While the temperature dependence of NCRI is well reproduced, the magnitude of NCRI is highly sample-dependent. In the experiment of Kim and Chan where the samples were confined within an annulus with a width of ~1 mm, NCRI was found to vary between 0.5% and 1.4% [2]. Many of the TO measurements were on bulk solid $^4$He grown in cells with cylindrical sample spaces with typical dimensions of ~1 cm [17]. The NCRI fraction (NCRIF) in these cylindrical samples varies between 0.015% and ~1%. In some samples the NCRIF can be reduced, sometimes dramatically, by thermal annealing [17,18]. The annealing effect and the variation in the NCRIF support the interpretation that various forms of disorder in the solid (grain boundaries, dislocation lines, glassy phases [19-22]) are either responsible for, or at least enhance the NCRI signal. Nevertheless, experiments on solid helium confined within disordered porous media such as Vycor glass [1] and porous gold [23] (with confining dimensions of 7 nm and 500 nm respectively) also found NCRI only on the order of 1-2%. A recent attempt to induce disorder in the helium by growing the solid inside of 95% porous silica aerogel did not lead to an enhanced NCRIF [24]. Rittner and Reppy (RR) carried out a series of TO measurements with the solid samples confined to thin annuli with 300, 150 and 74 μm gaps. Remarkably, they observed NCRIF as high as ~20% [18,25]. The results of RR suggest that there is an optimal surface to volume ratio, S/V, of the solid helium sample that is on the order of 100 to 1000 cm$^{-1}$ (or with confining dimension of ~150 μm) that is particularly conducive for the appearance of large NCRI. This extraordinary large NCRI is one of the most intriguing and puzzling results in the subject of superflow in solid helium.

In this paper we report a TO study of solid helium with confining dimension and S/V ratio similar to RR (~400 cm$^{-1}$). The internal geometry of the cell is similar to the original Andronikashvili liquid helium experiment [26]. The helium is confined to the space between thinly separated discs (150 μm gap). The oscillation axis is perpendicular to the plane of the thin discs. In contrast to RR, the NCRI observed in our sandwiched samples is less than 1%, in line with other TO results.

Previously, high precision heat capacity experiments on bulk solid $^4$He samples (using silicon calorimeters with a volume of ~ 1cc) found a peak in addition to the $T^3$ Debye contribution [4,27]. The temperature of this peak appears to be correlated with the onset of NCRI, both the position and the height of the peak show the same trend as NCRI with regard to quality of the sample. Specifically, a smaller peak at lower temperature is found for samples of higher crystalline quality.

To see if the confinement has any effect on the heat capacity signals, we also carried out heat capacity measurements in a calorimeter with an internal geometry similar to the TO cell. In the current experiment, the calorimeter body and the discs creating the confined geometry are made of sapphire. Similar to silicon, sapphire also has small heat capacity and high thermal conductivity at low temperatures. The change in materials was precipitated by the unfortunate characteristic of silicon that it cleaves easily along the crystal planes, making it a less than ideal pressure vessel. The background heat capacity of the empty sapphire calorimeter (including the discs and the epoxy gluing the cell together) is 10 times smaller than the helium signal and has been subtracted from the data shown in this paper.

The solid samples studied in both the TO and heat capacity measurements were made from standard ultra high purity (UHP) helium gas (with ~0.3 ppm of $^3$He) and grown using the blocked capillary (BC) method. In this method, after a high pressure liquid is introduced into the sample cell, the inlet capillary is blocked by a solid plug.

The pressure then drops as the sample is cooled along the liquid-solid coexistence curve until the solidification is complete.

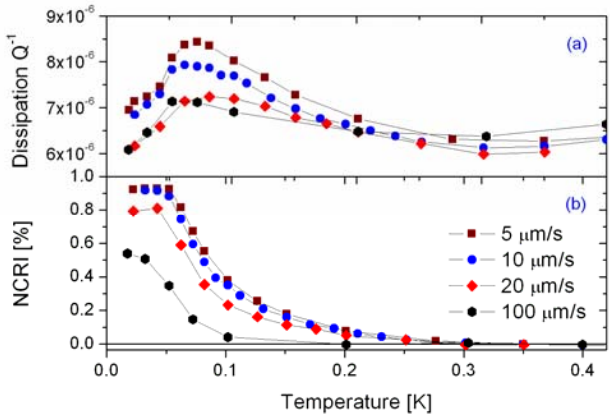

FIG. 1. NCRI and dissipation data for a 40 bar solid (sample A). The oscillator and thin discs creating the confined geometry are made of Be-Cu. The solid helium sample is composed of a stack of 19 thin annular discs (each with OD, ID and thickness of 12.5 mm, 6.1 mm and 150 μm respectively). The empty cell resonant period of this oscillator is 1.350575 ms. Filling the cell with a 40 bar sample increases the period by 1240 ns.

NCRI and dissipation data for sample A, a 40 bar solid are shown in Fig. 1. The liquid-solid conversion time in growing this sample is ~1 h. The data exhibit all of the characteristic features of NCRI. The onset of NCRI becomes resolvable below ~0.3 K. Accompanying the drop in period is a peak in the dissipation of the oscillator. The NCRI fraction is found to saturate at almost 1% in the low temperature limit. When the oscillation speed is increased beyond the critical velocity (~10 μm/s), a diminution in both NCRI and the onset temperature are seen. Thermal history dependence is also observed at higher drive (~100 μm) [28, 29].

Figure 2 shows the NCRI signal, measured at low oscillation speed, of three other solid samples. Sample D was grown (i.e. liquid to solid conversion time) in 15 minutes, the fastest rate possible given the total mass of the sample and how quickly the heat of fusion can be drained through the torsion rod, the 'bottle neck' of thermal conduction between the torsion bob and the mixing chamber. The rest of the samples were grown in ~1 h. Figure 2 shows that changing the growth rate from 15 to 60 minutes has no noticeable effect on the magnitude of NCRI. However, the high temperature tail of the NCRI signal is more pronounced for sample D, this is consistent with other lower quality solid samples [17]. Sample C, grown using the same procedure as samples A and B, shows a 0.6% NCRI instead of 0.9%. Sample C was subsequently warmed up to the melting curve then cooled down gradually over ~3 hours. The NCRI of the post-annealed sample is significantly smaller than before annealing. We note that the annealed sample has a higher onset temperature in contradiction to the trend shown in Ref. 17. The reason for the discrepancy is not clear.

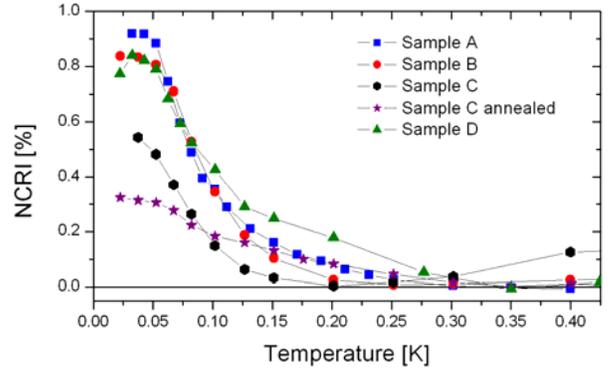

FIG. 2. NCRI data for four solid samples. Samples A, B and C are blocked capillary samples grown in ~1 hour with melting pressures of 40 bar, 45 bar and 39 bar respectively. Sample C was then thermal cycled over the course of 3 hours. Sample D (a 37 bar solid) was grown as quickly as possible, solidifying in 15 minutes. All data are cooling scans at low drive (~10 μm/s).

The specific heats of two samples grown between stacked sapphire discs are shown in Fig. 3(a). The growth times are 4 h and 20 h, respectively. For the calorimeter, the limiting factor in draining the heat of fusion is the weak thermal link (a thin copper wire) between the calorimeter and the mixing chamber. The fastest solidification time was 4 hours for this calorimeter. Longer growth periods were accomplished by cooling down the mixing chamber very slowly. For comparison, the specific heat of two bulk BC samples with the same growth times are also shown in Fig. 3. Below 0.15 K a deviation from the Debye-$T^3$ law is observed. When the $T^3$ term is subtracted, the excess specific heat takes the form of a peak for each of the four samples as shown in Fig. 3(b). Consistent with prior HC results [4,27], the data show no signs of a term that scales linearly with temperature, a characteristic signature of glassy behavior. As noted above, the peak height and peak temperature decrease with increasing sample quality for bulk samples. Such a dependence on growth time is not observed for the confined sample. In addition, the peak heights of these samples are considerable smaller and located at lower temperature than either of the two bulk samples. These findings suggest that crystals grown in this confined geometry possess consistently high crystallinity for reasons we do not fully understand. Perhaps the high thermal conductivity and the particular surface finish of the sapphire disks provide a favorable nucleation surface for the growth of solid helium.

Since the confining dimension and the S/V ratio of the stacked cells of our TO and HC experiments are similar to that of RR, how can one understand the contrasting results? The annular vs. planar confinement could be important. However, this is unlikely since the NCRIF does not appear to be directly correlated to bulk vs. annular geometries.

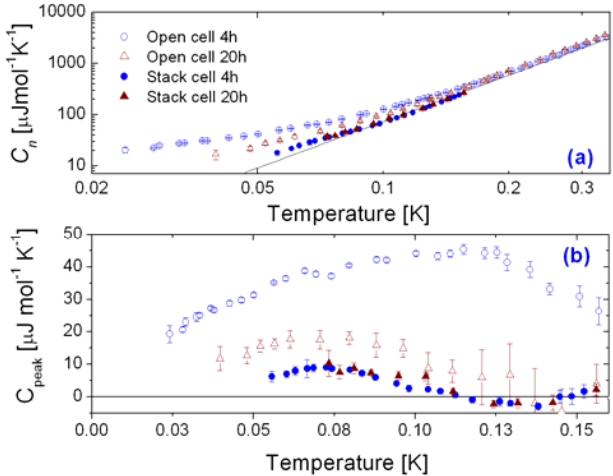

FIG. 3: Top panel (a) shows the heat capacity data for samples with different growth times for two different calorimeters. The open cell data is taken from Ref. [4,27]. Figure 3(b) shows the specific heat peak as a function of temperature with the phonon $T^3$ term subtracted (please note that the scales are different for the two panels). The pressures for these samples are 33±1 bar. The solid helium sample consists of a stack of 24 solid helium discs with a thickness of 150 μm and diameter of 15 mm. The volume of the cell is 0.71 cc.

One explanation for the unusually large NCRI seen by RR is related to sample growth time and the fact that the volumes of their samples are significantly smaller than all other TO samples, including samples embedded in porous media. The heat of fusion for helium is very large in comparison to the thermal mass of the torsion bob. The heat of fusion for $^4$He at 2.5 K is ~0.3 J/cc [30]. In comparison, the amount of energy removed when cooling either Be-Cu or solid $^4$He from 2.5 K to 1 K is ~10 times less [30,31]. The typical cylindrical cell has ~1 cc of both metal and $^4$He. Therefore the shortest time it takes to grow a solid sample from liquid in a TO is typically limited by how quickly the heat of fusion is drained through the torsion rod to the mixing chamber. This characteristic time is roughly given by $LV/G_T\Delta T$, where L is the latent heat of fusion, V, the sample volume, $\Delta T$, the temperature gradient along the torsion rod during liquid-solid conversion, and $G_T$ is the thermal conductance of the torsion rod given by $kA/l$, where k is the thermal conductivity of the rod, A is the cross-sectional area, and l, the length. It has not been a priority and it is not easy to measure the precise time it takes to solidify a liquid sample in each of the many TO experiments conducted in the various laboratories. Similarly, $\Delta T$ during solidification depends on the exact cooling protocol used by the different experimental groups. For the numerous BC samples made at Penn State with BeCu torsion rods, $\Delta T$ typically varies between 0.5 K and 1 K. In Table I, we tabulate V, $R_S$, and NCRIF of all the TO experiments to date. Since the NCRIF can be reduced by annealing and careful sample growth procedures, the maximum reported NCRIF values from experiments using UHP $^4$He gas are listed. The relative solidification times for the different TO's can be roughly compared by calculating the quantity we call the quenching factor, $QF=G_T/V$. A larger QF indicates a shorter time required to grow the sample. Figure 4 shows the NCRI as a function of the QF for each experiment. In spite of the different cooling procedure used by the different experimental groups in growing the crystals, there is clearly a trend of NCRIF increasing with QF regardless of the sample confinement (i.e. bulk, thin or thick annulus, or porous media). Two experiments using AgCu for the torsion rod have shown small NCRI and do not fit into this trend. Because of the high thermal conductivity of AgCu, our assumption that the torsion rod is the limiting factor in freezing the sample is not valid and the solidification time is determined by the cooling rate of the cryostat.

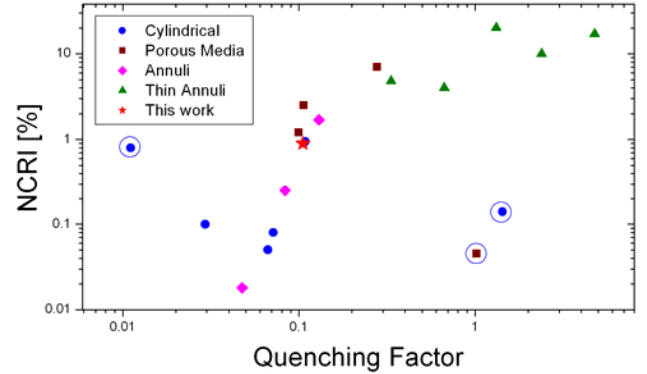

FIG. 4. NCRI increases with QF (i.e. faster solidification time). The circled outliers on the right are the two AgCu cells (as explained in the text) and the one on the left is the square cell from RR [10].

Figure 4 indicates that the large NCRI observed by RR and others [15,35] may be due to this quenching factor which allows for quick solidification and not the result of confinement at any special length scale. RR reported the time it took to cool the sample in their thin annulus cell [18] from liquid down to below 1 K is about 90 s. The minimum time required for solidifying blocked capillary samples in the other experiments is typically 15 minutes or greater. It appears in order for quench cooling to produce sufficient disorder to give rise to a large (~10%) NCRI, the liquid sample must be frozen on the time scale of ~1 minute or less.

As noted above, dislocation network, grain boundaries and glassy phase are the likely sources of disorder that enhance NCRI. However, it is not possible to reconcile 10% NCRIF with models of superfluidity based on networks of dislocation and grain boundaries. This leaves the quenched-in glassy phase in solid helium as the likely origin for the large NCRI observed by RR. Indeed Hunt and collaborators at Cornell recently reported a large (4.5%) NCRIF in their TO with a 100 μm annular gap and they also observe evidence of glassy behavior [15]. Time-dependent phenomenon such as hysteresis and long time scale relaxation processes were also reported by Clark et al. and Aoki et al. [28,29]. It will be interesting to look for structural evidence of such a highly disorder solid (e.g. a glassy state) by means of x-ray or neutron scattering.

| Geometry | Ref. | V cm$^3$ | $G_T$ cm (x$10^{-2}$) | NCRIF % |
|---|---|---|---|---|
| Bulk Samples | 17 BeCu | 0.32 | 3.45 | 0.95 |
| | 17 AgCu | 0.48 | 69 | 0.14 |
| | 11 | 0.62 | 1.84 | 0.1 |
| | 12 | 0.40 | 2.6 | 0.05 |
| | 34 | .314 | 2.2 | .08 |
| | 10 | 1.4 | 1.55 | 0.8 |
| Porous Media | 1 | 0.212 | 2.06 | 2.5 |
| | 23 | 0.055 | 0.528 | 1.2 |
| | 35 | 0.064 | 1.8 | 7 |
| | 24 AgCu | 0.503 | 51 | 0.045 |
| Annular | 2 | 0.16 | 2.06 | 1.7 |
| | 36 | 0.23 | 1.84 | 0.25 |
| | 37 | .225 | 1.06 | 0.018 |
| 300 μm | 18 | 0.19 | 12.7 | 4 |
| 150 μm | 18 | 0.096 | 12.7 | 20 |
| 150 μm | 25 Al | 0.090 | 21.2 | 10 |
| 76 μm | 25 Al | 0.044 | 21.2 | 17 |
| 100 μm | 15 | 8.5x$10^{-3}$ | 0.28 | 4.8 |
| This work | | 0.38 | 4.0 | 0.9 |

TAB. I. The values for V, $G_T$, and NCRIF are listed. For simplicity we normalize k for the different torsion rod materials at 4 K to the value for BeCu. Therefore, we use k=1 for BeCu and k=2 for Al and k=20 for AgCu [32,33].

We acknowledge informative discussions with A. S. Rittner and J. D. Reppy. We also thank S. Davis, E. Kim, H. Kojima, M. Kubota, and K. Shirahama for providing details of their experiments. This work is supported by NSF under grant DMR 0706339.